\documentclass[preprint,showpacs,amssymb,aps,prd]{revtex4}

\begin{document}

%

\title{Semiclassical Methods in Curved Spacetime 
\\
and Black Hole Thermodynamics}

\author{Horacio E. Camblong$^{1}$
and
Carlos R. Ord\'{o}\~{n}ez$^{2,3}$}

\affiliation{
$^{1}$
Department of Physics, University of San Francisco, San
Francisco, California 94117-1080
\\
$^{2}$ Department of Physics, University of Houston, Houston,
Texas 77204-5506
\\
$^{3}$
World Laboratory Center for Pan-American Collaboration in Science and
Technology,
\\
University of Houston Center, Houston, Texas 77204-5506
}

\begin{abstract}
Improved semiclassical techniques are developed
and applied to a treatment of a real scalar field in a $D$-dimensional 
gravitational background. This analysis, leading to a derivation of the 
thermodynamics of black holes, is based on the simultaneous use of
(i) a near-horizon description of the scalar field in terms of conformal 
quantum mechanics; (ii) a novel generalized WKB framework; and (iii) 
curved-spacetime phase-space methods. In addition, this improved 
semiclassical approach is shown to be asymptotically exact in the 
presence of hierarchical expansions of a near-horizon type. Most 
importantly, this analysis further supports the claim that the 
thermodynamics of black holes is induced by their near-horizon 
conformal invariance.
\end{abstract}


\maketitle

\section{Introduction}
\label{sec:introduction}

 The fundamental concepts of black hole thermodynamics have been 
confirmed within several frameworks since the 1970s~\cite{BH_thermo_reviews},
including in string theory~\cite{BH_string} and 
loop quantum gravity~\cite{BH_loop}.
In particular, the Bekenstein-Hawking entropy
$S_{\rm BH}$~\cite{Bekenstein_entropy} 
and the Hawking effect~\cite{Hawking_temperature} 
suggest that the  horizon plays a fundamental role in black hole 
thermodynamics~\cite{thooft:85,frolov}, an idea 
that has been emphasized in recent 
approaches~\cite{BH_thermo_reviews,padmanabhan_review_04}
and generalized to the holographic principle~\cite{holographic,AdS/CFT}.
The connections between the horizon quantum features and the thermodynamics
include the existence of a near-horizon conformal 
symmetry~\cite{strominger:98,gov:BH_states,gupta:BH,carlip:near_horizon,solodukhin:99,BH_superconformal}. 
In Ref.~\cite{BH_thermo_CQM} we have discussed
 the emergence of this thermodynamic behavior,
within a {\em semiclassical approximation\/} with
the following building blocks:
(i) the near-horizon conformal symmetry;
(ii) the competition of the field angular-momentum degrees of freedom;
and (iii) the singular dynamics of conformal 
quantum mechanics~\cite{jackiw,fubini_beyond}.
Given the
 {\em singularity of the conformal potential\/},
these ingredients suggest the questions:

(i) Is the use of semiclassical techniques justified within 
 conformal quantum mechanics~\cite{camblong}?

(ii) Is there a preferred order
for the angular-momentum expansion {\em vis-\`{a}-vis\/}
the radial conformal analysis?

In this paper, 
we give an affirmative answer to the first question through
an improved semiclassical method, and show that 
 the stage at which the field
angular-momentum expansion is introduced is
immaterial; thus,
our framework justifies the use of these concepts.

 In Sec.~\ref{sec:QFT_eqs}
we survey the scalar field 
equations, including their near-horizon properties.
In Sec.~\ref{sec:n-h_generalized_WKB}
we develop a generalized version of 
the semiclassical WKB method 
in the presence of a hierarchical expansion---the near-horizon expansion
being a particular case.
In Sec.~\ref{sec:phase_space_semiclassical_methods}
we use phase-space methods in curved spacetime
to derive the spectral function needed for the thermodynamics.
Finally,
in Sec.~\ref{sec:BHEntropy_conclusions},
we discuss and critically reexamine this framework.

\section{Field Equations}
\label{sec:QFT_eqs}

We will consider
a real scalar field $\Phi$, with mass $m$,
in a $D$-dimensional spacetime ($D \geq 4$),
defined through its action
(with the metric conventions of Ref.~\cite{misner_thorne_wheeler})
\begin{equation}
S
=
-
\frac{1}{2}
\int
d^{D} x
\,
\sqrt{-g}
\,
\left[
g^{\mu \nu}
\,
\nabla_{\mu} \Phi
\, 
\nabla_{\nu} \Phi
+ 
m^{2} \Phi^{2}
+  \xi R \Phi^{2}
\right]
\; ,
\label{eq:scalar_action}
\end{equation}
where $\xi$ is
its coupling to the curvature scalar $R$,
and the static spacetime metric  is 
\begin{equation}
 ds^{2}
=
g_{00} ( \vec{x})
\,  dt^{2}
+
\gamma_{ij} ( \vec{x})
 d x^{i} dx^{j}
= 
- f (r) \,  dt^{2}
+
\left[ f(r) \right]^{-1} \, dr^{2}
+ r^{2} \,
 d \Omega^{2}_{(D-2)}
\; , 
\label{eq:RN_metric}
\end{equation}
where
$
d \Omega^{2}_{(D-2)}
$
is the metric on $S^{D-2}$.
The derivation of the thermodynamics
requires counting the field modes for
the spectral number function $N(\omega)$ leading to the
entropy; thus,
this procedure is based 
on the combinatorics of the modes of the
Euler-Lagrange equation 
\begin{equation}
\Box 
\,
\Phi
- 
\left(
m^{2} 
+ 
\xi R
\right)
\Phi
= 0
\; 
\label{eq:Klein_Gordon_basic}
\end{equation}
satisfied by $\Phi$ from the action~(\ref{eq:scalar_action}).
The  quantum field operator
can be expanded as
\begin{equation}
 \Phi(t, \vec{x} )
= \sum_{s}
\left[
 	a_{s}
\,
\phi_{s} (\vec{x} )
\,
e^{-i\omega_{s} t }
 	+ a^{\dagger}_{s}
\,
\phi^{*}_{s} (\vec{x} )
\,
e^{i\omega_{s}t}
 	\right]
\; ,
\label{eq:field_Fourier_expansion}
\end{equation}
where $a^{\dagger}_{s}$ and $a_{s}$ are the creation and annihilation 
operators, and 
$\phi_{s} (\vec{x} )$ is a complete set of orthonormal 
functions; 
the 
separation of the time coordinate
of Eq.~(\ref{eq:field_Fourier_expansion}) turns
 Eq.~(\ref{eq:Klein_Gordon_basic})
into
\begin{equation}
\check{\Delta}_{(\gamma)} \phi
+
\gamma^{ij}
\partial_{i} 
\left( 
\ln \sqrt{ |g_{00}|} 
\right)
\partial_{j} \phi
+
\mathfrak{I}_{(0)} (r; \omega)
\,
\phi 
=
0
\; ,
\label{eq:Klein_Gordon_reduced_to_spatial}
\end{equation}
where
$\check{\Delta}_{(\gamma)} $ 
is the Laplace-Beltrami operator
with respect to the spatial metric $\gamma_{ij} (\vec{x}) $
and
\begin{equation}
\mathfrak{I}_{(0)} (r; \omega)
=
\frac{\omega^{2}}{f (r) }
-\left(
m^{2}
+
\xi  R
\right)
\; .
\label{eq:original_terms_I0}
\end{equation}

An important ingredient for our thermodynamic analysis
is the {\em near-horizon expansion\/}, which is defined
with respect to the coordinate
\begin{equation}
x= r -r_{+}
\; ,
\label{eq:n-h_coordinate}
\end{equation}
where 
$r=r_{+}$ selects the event horizon
 ${\mathcal H}$ from the largest root 
of the scale-factor equation $f(r)=0$
(excluding cosmological horizons).
Given a quantity $Q$, for a leading order $s$,
it will prove useful to use the notation
$
Q(r) 
\stackrel{(\mathcal H)}{\sim} 
Q^{(s)}_{+}  \, x^{s}/\Gamma (s+1)
$,
which amounts to performing a Laurent expansion;
in the case of a Taylor series expansion,
$Q^{(s)}_{+}   $ stands for the $s\/$th-order derivative of $Q(r)$ at
${\mathcal H}$.
In particular, we will consider the parameter
\begin{equation}
f'_{+}
\equiv f'(r_{+})
\; ,
\end{equation}
with
$
f'_{+}
\neq 0 
$
for the 
the {\em nonextremal\/} case; this
 entails the leading-order expansion
$f(r)
\stackrel{(\mathcal H)}{\sim}
f'_{+}  \, x
\left[ 1 + O(x) \right]
$,
and its corresponding higher orders
$f'(r)
 \stackrel{(\mathcal H)}{\sim}
f'_{+}
$ and
$
f''(r)
 \stackrel{(\mathcal H)}{\sim}
f''_{+} $.

Several paths can be taken to describe 
the relevant physics of the modes. The first one
involves the Liouville transformation~\cite{forsyth:Liouville}
$
\phi (\vec{ x} )
=
| g_{00}|^{-1/4}
\,
\Psi ( \vec{x})
$,
so that Eq.~(\ref{eq:Klein_Gordon_reduced_to_spatial}) becomes
\begin{equation}
\check{\Delta}_{(\gamma)} \Psi
+
\mathfrak{I} (r;\omega)
\,
\Psi
=
0
\; ,
\label{eq:Klein_Gordon_normal_spatial}
\end{equation}
whose normal form
involves the  extra terms
\begin{equation}
\mathfrak{I}_{(1)} (r)
= 
f(r)
\left\{
\frac{1}{16} \, \left[ \frac{f'(r)}{f(r)} \right]^{2}
-
\frac{D-2}{4 \, r} \, \frac{f'(r)}{f (r)}
-
\frac{1}{4} \, \frac{f''(r)}{f(r)}
\right\}
\;  ,
\label{eq:extra_terms_I1}
\end{equation}
in addition to the original function
$\mathfrak{I}_{(0)} (r;\omega)$ of Eq.~(\ref{eq:original_terms_I0}),
with
\begin{equation}
\mathfrak{I} (r;\omega)
 = 
\mathfrak{I}_{(0)} (r;\omega)
+
\mathfrak{I}_{(1)} (r)
\;  .
\label{eq:Klein_Gordon_normal_QMreduced_invariant}
\end{equation}

Finally, the near-horizon expansion of 
Eq.~(\ref{eq:Klein_Gordon_normal_spatial}) involves the 
conformally-symmetric terms
\begin{equation}
\mathfrak{I} (r;\omega) 
\stackrel{(\mathcal H)}{\sim}
f'_{+}
\,
\left(
\Theta^{2}
+
\frac{1}{16}  
\right)
\,
\frac{1}{ x }
\, 
\left[ 1 + O(x) \right]
\; ,
\label{eq:Klein_Gordon_normal_QMreduced_invariant_NH}
\end{equation}
with the same scaling as the
Laplace-Beltrami operator and
 characterized
by the parameter
\begin{equation}
\Theta
= 
\frac{\omega}{  f'_{+} } 
\;  .
\label{eq:conformal_parameter_theta}
\end{equation}

The second path consists of introducing the 
spherical symmetry of the metric~(\ref{eq:RN_metric})
directly from the outset,
so that Eq.~(\ref{eq:Klein_Gordon_reduced_to_spatial}) turns into
\begin{equation}
 f(r) \, \phi''
+ \left[ 
f'(r) + 
(D-2) \, \frac{ f(r) }{r} 
\right]
\phi'
- \frac{1}{r^{2}} 
\,
\hat{\ell}^{2} 
\phi
+ 
\mathfrak{I}_{(0)} (r;\omega)
\,
\phi 
= 
0
\; ,
\label{eq:Klein_Gordon_radial_metric}
\end{equation}
where  $- \hat{\ell}^{2}  = -\hat{\ell}^{a} \hat{\ell}_{a}$
stands for the Laplacian on $S^{D-2}$,
with its spherical-harmonic eigenfunctions
$Y_{l m} ( \Omega )$.
In addition, applying the Liouville 
transformation~\cite{forsyth:Liouville}
\begin{equation}
\phi_{s} (\vec{x} )
\equiv
\phi_{nl m} (r,\Omega )
= 
Y_{l m} (  \Omega )
\,
\chi (r)
\,
u_{nl} (r)
\; 
\end{equation}
[where $s=(n,l,m)$],
with
$
\chi (r)
=
[f(r)]^{-1/2} 
\,
r^{-(D-2)/2}
$,
the radial equation becomes
\begin{equation}
u''(r) 
+
\mathfrak{I} (r; \omega, \alpha_{l,D}  ) 
\,
u (r) 
=0
\;  ,
\label{eq:Klein_Gordon_normal_radial}
\end{equation}
in which
\begin{eqnarray}
\!  \! \!  \!  \!  \! 
\mathfrak{I}   
(r; \omega, \alpha_{l,D}  ) 
& = &
\mathcal{I} (r;\omega) 
-
\frac{1}{f (r)} 
\,
\frac{ l \left( l + D-3 \right)  }{r^{2}} 
\label{eq:I_term-vs-J_term}
\\
\!  \! \!  \! \!  \! 
& = &
\frac{ \mathfrak{I}_{(0)} (r;\omega) }{ f(r)}
+ \left\{
\left[ 
\frac{1}{f(r)} - 1 \right]
\, 
\nu^{2} + \frac{1}{4}
\right\}
\,
\frac{1}{r^{2}} 
+
R_{rr}
+
\frac{1}{4}
\left[
\frac{f'(r)}{ f(r)}
\right]^{2}
-
\frac{1}{f (r)} 
\,
\frac{ \alpha_{l,D} }{r^{2}} 
\;  
\label{eq:Klein_Gordon_normal_radial_invariant}
\end{eqnarray}
includes the terms
$\mathcal{I} (r;\omega)
 $
associated with the radial Liouville transformation,
while 
\begin{equation}
\alpha_{l,D} 
=
l(l+D-3) 
+ \nu^{2} = 
\left(
l + \frac{D-3}{2} 
\right)^{2}
\; 
\label{eq:alpha_angular_momentum}
\end{equation}
is the angular-momentum coupling.
In Eqs.~(\ref{eq:Klein_Gordon_normal_radial_invariant})
and (\ref{eq:alpha_angular_momentum}),
$
\nu = (D-3)/2$
and
\begin{equation}
R_{rr}
=
-
\frac{1}{2}
\frac{f''(r)}{ f(r)}
-
\frac{(D-2)}{ 2 \, r} \, \frac{ f'(r) }{  f(r) } 
\;  
\label{eq:Ricci_rr}
\end{equation}
is the radial component of the Ricci tensor
for the metric~(\ref{eq:RN_metric}).

The most important property of 
Eq.~(\ref{eq:Klein_Gordon_normal_radial_invariant})
 is that its near-horizon expansion,
\begin{equation}
\mathcal{I} (r;\omega) 
\stackrel{(\mathcal H)}{\sim}
\left(
\Theta^{2}
+
\frac{1}{4}  
\right)
\,
\frac{1}{ x^{2} }
\, 
\left[ 1 + O(x) \right]
\; ,
\label{eq:scalar_field_in_BH_background_6_invariant_NH}
\end{equation}
is {\em conformal\/}
because $\mathcal{I} (r;\omega) $ has the same
scale dimension as the second-order derivative
in Eq.~(\ref{eq:Klein_Gordon_normal_radial}).
Comparison of Eqs.~(\ref{eq:Klein_Gordon_normal_QMreduced_invariant_NH})
and (\ref{eq:scalar_field_in_BH_background_6_invariant_NH})
shows that:
(i) the scale dimension is changed from $1/x$ to $1/x^{2}$;
and (ii) 
the numerical term has changed from 1/16 to 1/4.
The first point is due to a rearrangement of factors:
Eqs.~(\ref{eq:Klein_Gordon_normal_QMreduced_invariant_NH})
and (\ref{eq:scalar_field_in_BH_background_6_invariant_NH})
 describe the same physics within different 
{\em coordinate representations\/} of conformal quantum mechanics.
The second, subtler point is crucial
for the counting of modes, as will be
seen in Secs.~\ref{sec:n-h_generalized_WKB}
and \ref{sec:phase_space_semiclassical_methods}.

Finally,
including the angular momentum, 
the near-horizon expansion of 
Eq.~(\ref{eq:I_term-vs-J_term})
is
\begin{equation}
\mathfrak{I}   (r; \omega, \alpha_{l,D}  ) 
\stackrel{(\mathcal H)}{\sim}
\left\{
\left[
\frac{ \omega^{2} }{(f'_{+})^{2}}
+ \frac{1}{4} 
\right]
\,
\frac{1}{ x^{2} }
-
\frac{ \alpha_{l,D}  }{
f'_{+} \, r_{+}^{2}}
\frac{1}{x}
\right\}
\, 
\left[ 1 + O(x) \right]
\; ,
\label{eq:scalar_field_in_BH_background_5_invariant_NH}
\end{equation}
which displays the properties:
(i) the leading term is the
strong-coupling potential
\begin{equation}
V_{\rm eff} (x) 
\stackrel{(\mathcal H)}{\sim}
- 
 \left(  
\Theta^{2} + \frac{1}{4}
 \right) 
\,
\frac{1}{x^{2}}
\, 
\left[ 1 + O(x) \right]
\; 
\label{eq:conformal_interaction}
\end{equation}  
of conformal quantum 
mechanics~\cite{BH_thermo_CQM};
(ii)
the angular-momentum term
is still required for the correct statistical counting of modes leading 
to the thermodynamics~\cite{BH_thermo_CQM}.

\section{Near-Horizon Generalized WKB Framework}
\label{sec:n-h_generalized_WKB}

We will consider the effective 
problem obtained after separation of the time coordinate,
which consists of a $d$-dimensional
equation (with spacetime dimensionality $D=d+1$)
\begin{equation}
\check{\Delta}_{(\gamma)} 
\Psi
+
\mathfrak{I} (\vec{x} )
\,
\Psi
=
0
\;  .
\label{eq:Schrodinger_equation}
\end{equation}

\subsection{Covariant WKB Method: General Formulation}
\label{sec:WKB}

Our goal is to select a WKB wave vector that would reproduce 
the original equation~(\ref{eq:Schrodinger_equation})
as closely as possible.
Without loss of generality, 
one can start from a  WKB-type solution 
\begin{equation}
 \Psi_{{\rm WKB}} ( \vec{x} ) 
=
A  ( \vec{x} ) 
 \exp \left[
 i \int^{   \vec{x} } 
k_{j} (  \vec{x} \,  ') \, d   x'^{j}
\right]
\; ,
\label{eq:covariant_WKB_wf}
\end{equation}
in which the wave number $k_{j} ( \vec{x} ) $
and amplitude $A  ( \vec{x} )$ are real.
This is known to be a first-order approximation 
in an expansion with respect to the ``small'' parameter
$\hbar$,
but may fail to be an exact solution
of the  problem~(\ref{eq:Schrodinger_equation}).
However, defining
\begin{equation}
\tilde{ \mathfrak{I} } (\vec{x} ) =
\| \vec{k}  (\vec{x}) \|^{2}
\equiv 
\gamma^{jh} ( \vec{x} )  
\,
k_{j} ( \vec{x}  ) k_{h}  ( \vec{x} )
\; ,
\label{eq:eta_definition}
\end{equation}
the wave function~(\ref{eq:covariant_WKB_wf}) satisfies
the {\em exact\/} equation
\begin{equation}
\check{\Delta}_{(\gamma)} 
\Psi_{\rm WKB}
+
\left[ 
\tilde{ \mathfrak{I} } (\vec{x} ) 
- Q  ( \vec{x} ) 
\right]
\,
\Psi_{\rm WKB}
=
0
\;  ,
\label{eq:exact_covariant_WKB_equation}
\end{equation}  
which follows by enforcing the  conservation
of the ``effective probability current'' $j_{h}= A^{2} k_{h} $:
\begin{equation}
\nabla_{j} \left[\gamma^{jh} 
A^{2} k_{h} 
\right]
\equiv
\frac{1}{ \sqrt{\gamma} }
\,
\partial_{j} \left[\gamma^{jh} \sqrt{\gamma}
\,
A^{2} k_{h} 
\right]
= 0
\; ,
\label{eq:probability_conservation}
\end{equation}
thus suppressing the terms associated with imaginary coefficients,
and  leading to 
\begin{equation}
Q  ( \vec{x} ) 
=
\frac{\check{\Delta}_{(\gamma)} 
A ( \vec{x} )}{A ( \vec{x} ) }
\; .
\label{eq:covariant_WKB_error}
\end{equation}

Traditionally,
 the function $Q  ( \vec{x} ) $
in Eq.~(\ref{eq:covariant_WKB_error}) is viewed as the ``error''
in approximating $\Psi  ( \vec{x} ) $
with $\Psi_{\rm WKB}  ( \vec{x} ) $, with applicability 
limited by $|Q  ( \vec{x} ) | \ll \| \vec{k}  (\vec{x}) \|^{2}   $.
However, for the near behavior $x \sim 0$,
a modified WKB approach, in the style first proposed by 
Langer~\cite{langer}, may be needed.
We will consider a generalized covariant 
scheme that expands the range of applications and
permits a treatment of the coordinate singularity.
In this proposal, the additional term
$Q  ( \vec{x} ) $ in Eq.~(\ref{eq:exact_covariant_WKB_equation}) is 
absorbed by the original function 
$\mathfrak{I} (\vec{x} ) $ in Eq.~(\ref{eq:Schrodinger_equation}),
in such a way that
$\Psi_{\rm WKB}  ( \vec{x} ) 
=
\Psi  ( \vec{x} ) $;
thus, from Eqs.~(\ref{eq:Schrodinger_equation}) and
(\ref{eq:exact_covariant_WKB_equation}),
 it follows that 
\begin{equation}
\tilde{ \mathfrak{I} } (\vec{x} ) 
= 
\mathfrak{I}  ( \vec{x} )  
+ 
Q [ \tilde{ \mathfrak{I} } ] ( \vec{x} )
\;   ,
\label{eq:WKB_quantum_potential_correction}
\end{equation}
which is an auxiliary equation,
where $Q ( \vec{x} ) $ depends on the unknown 
 $\tilde{ \mathfrak{I} } ( \vec{x} ) $ and its derivatives.
Thus,
the improved WKB method amounts to
the replacement
$
\mathfrak{I} (\vec{x} ) 
\rightarrow
\tilde{\mathfrak{I}} ( \vec{x} ) 
$, where
the subtraction of the ``quantum potential'' $ Q ( \vec{x} ) $
generates an {\em effective potential\/}
$-\tilde{\mathfrak{I}} ( \vec{x} )$
that captures the relevant
physics.
In this viewpoint,
Eqs.~(\ref{eq:eta_definition})
and
(\ref{eq:probability_conservation})--(\ref{eq:WKB_quantum_potential_correction})
constitute a set of coupled partial differential equations;
even though an  exact solution 
to this combined system 
is not generally available,
a systematic approximation scheme
can be developed as follows.
Specifically, Eq.~(\ref{eq:WKB_quantum_potential_correction}) 
is taken as the starting point
of a successive-approximation scheme
\begin{equation}
\tilde{ \mathfrak{I} }^{(n)} 
( \vec{x} ) 
=
\mathfrak{I}  ( \vec{x} ) 
 + Q [ \tilde{ \mathfrak{I} }^{(n-1)}  ]
( \vec{x} ) 
\;  ,
\label{eq:diff_eq_for_eta_nth-order}
\end{equation}
which begins at zeroth order ($n=0$) with the standard WKB approximation
\begin{equation}
\tilde{ \mathfrak{I} }^{(0)} ( \vec{x} ) 
=
\mathfrak{I}  ( \vec{x} ) 
\;  .
\label{eq:diff_eq_for_eta_0th-order}
\end{equation}
We now turn to the development
of a novel approximation framework,
which follows when this scheme 
is applied concurrently
with an expansion of the near-horizon type.

\subsection{Generalized WKB Framework
 in the Presence of a Hierarchical Expansion}
\label{sec:WKB_Expansion}

As discussed throughout this paper, the emergence of black hole thermodynamics
is governed by the near-horizon behavior of the metric~(\ref{eq:RN_metric}),
which can be displayed by means of an expansion with respect to the 
coordinate $x$ of Eq.~(\ref{eq:n-h_coordinate}).
The existence of an expansion of this kind furnishes a hierarchy, 
which organizes the relevant physics  with respect 
to powers of the variable $x$, starting with the dominant physics 
for the leading order.
Such a {\em hierarchical expansion\/} can be conveniently
applied concurrently with the (covariant) WKB approach of the 
previous subsection to provide a systematic modified WKB approach.
As we will show next, within the ensuing 
{\em hierarchical WKB framework,\/} the first-order 
approximation ($n=1$) in 
Eq.~(\ref{eq:diff_eq_for_eta_nth-order})
becomes {\em asymptotically exact\/}
with respect to $x \sim 0$, so that
\begin{equation}
\tilde{ \mathfrak{I} } ( \vec{x} ) 
\sim
\tilde{ \mathfrak{I} }^{(1)} ( \vec{x} ) 
=
\mathfrak{I} (\vec{x} ) 
 + Q [  \mathfrak{I}   ] ( \vec{x} ) 
\; ,
\label{eq:n-h_diff_eq_for_eta}
\end{equation}
where $\sim$ stands for the hierarchical expansion
 [with the near-horizon case being 
$\stackrel{(\mathcal H)}{\sim}$].

The dominant physics is described by the leading orders of the 
building blocks of Eq.~(\ref{eq:Schrodinger_equation}):
$\mathfrak{I} (x) $ and $\check{\Delta}_{(\gamma)} $.
In the hierarchical WKB framework,
the relevant expansion variable is $x$,
which we choose with dimensions of length.
Then, the {\em leading scale dimensions\/} of
$\mathfrak{I} (x) $, $\check{\Delta}_{(\gamma)}  $, and other variables
can be 
identified from the homogeneous degree of the
leading-order terms, under a rescaling 
$x \rightarrow \lambda \, x$.
Specifically, 
the dimension $[\check{\Delta}_{(\gamma)} ] = - p$ 
can extracted from
\begin{equation}
\frac{ \check{\Delta}_{(\gamma)} F (x)}{ F(x) }
\sim  \chi (s) \, x^{p}
\;  ,
\label{eq:Laplace_Beltrami_scale-dimension}
\end{equation}
while $[\mathfrak{I} (x) ] = - q$
is defined by 
\begin{equation}
\mathfrak{I} (x) \sim c \,  x^{q}
\;  .
\label{eq:potential_I_leading}
\end{equation}
In Eq.~(\ref{eq:Laplace_Beltrami_scale-dimension}),
 the test function $F(x)$ admits the expansion
$
F (x) \sim 
F^{(s)} \, x^{s}
/\Gamma ( s + 1 )  
$,
while $\chi (s)$ is a normalization factor that depends on
the dimension parameter $s$ associated with $F(x)$.
The scale dimension of $Q(x)$
can be determined 
from Eqs.~(\ref{eq:covariant_WKB_error})
and (\ref{eq:Laplace_Beltrami_scale-dimension})
(with $A \equiv F$),
\begin{equation}
Q (x) 
\sim  \chi (s) \, x^{p}
\;  ,
\label{eq:quantum_potential_leading}
\end{equation}
where the normalization function $\chi (s)$ is to be
computed from the specific expansion of the operator
$\check{\Delta}_{(\gamma)} $ with respect to $x$; thus,
the ``quantum potential'' $Q(x)$
has the same scale dimension, $-p$,
as the Laplace-Beltrami operator.
As a result,
Eq.~(\ref{eq:WKB_quantum_potential_correction})
defines the scale dimension 
of 
$\tilde{ \mathfrak{I} } (x)$ by selecting the leading order, i.e.,
$
[ \tilde{ \mathfrak{I} } (x) ] = - {\rm min} \{ p,q \}
$.

In addition to the scale dimension displayed in 
Eq.~(\ref{eq:quantum_potential_leading}), it is necessary to determine 
the normalization prefactor $\chi (s)$,
whose functional form can be computed from
the derivatives in Eq.~(\ref{eq:covariant_WKB_error}).
However, the actual value of $s$ requires the use
of the continuity condition~(\ref{eq:probability_conservation}),
combined with Eq.~(\ref{eq:eta_definition})
(see the near-horizon expansion  in the next subsection).

The nature of the expansion leads to three possible scenarios
 from a comparison of the 
scale dimensions of $\mathfrak{I}(x)$ and
$\check{\Delta}_{(\gamma)} $:
{\em regular case},
defined by $q>p$, so that the Laplace-Beltrami operator 
yields the dominant physics near $x \sim 0$;
{\em properly singular case},
defined by  $q<p$, so that $\mathfrak{I}(x)$ is dominant
as $x\sim 0$;
{\em marginally singular case}, 
defined by $q=p$, so that $\mathfrak{I}(x)$ and
the Laplace-Beltrami operator 
[along with the quantum potential $Q(x)$]
compete at the same order.
As for the solutions,
for the regular case,
they are of power-law  free-particle type as $x \sim 0$;
in addition,
for the singular cases $q \leq p$,
asymptotically exact WKB solutions can be found 
by:
\begin{enumerate}
\item
The {\em standard WKB method,\/} 
for the properly singular case.
In this method, the required effective potential
$-\tilde{\mathfrak{I}}  (\vec{x}) $  
only involves the term
$- \mathfrak{I}  (\vec{x})$,
with negligible 
$Q(\vec{x})$.
\item
The {\em improved WKB method,\/}
which applies to the marginally singular case.
In this method, the required effective potential
$- \tilde{\mathfrak{I}} (\vec{x}) $  is given 
from the rule~(\ref{eq:WKB_quantum_potential_correction})
or (\ref{eq:n-h_diff_eq_for_eta}).
\end{enumerate}

The latter, nontrivial case can be established
by going back to Eq.~(\ref{eq:diff_eq_for_eta_nth-order})
and verifying it becomes self-consistent at the $n=1$
level, in the form of Eq.~(\ref{eq:n-h_diff_eq_for_eta}).
Moreover, substituting Eqs.~(\ref{eq:potential_I_leading}) and
(\ref{eq:quantum_potential_leading})
in Eq.~(\ref{eq:n-h_diff_eq_for_eta}), and defining 
$c^{(*)}= - \chi (s)$, we see that 
\begin{equation}
\tilde{ \mathfrak{I} } (x)
\sim
\left[
c  - c^{(*)}
\right]
\, x^{p}
\; .
\label{eq:critical_coupling_subtraction}
\end{equation}
Thus, the nature of the modes 
changes around $c=  c^{(*)}$,
which plays the role of a critical coupling,
with $c$ selecting
either a singular (supercritical) or regular (subcritical) 
behavior. 

In conclusion, singular
quantum mechanics can be described with asymptotic
exactness by the improved WKB method, with
 modes having a semiclassical appearance
due to the singular term $\mathfrak{I}(x)$.
 However, in the marginally 
singular case, the competing ``potential'' $Q(x)$
generates the subtraction of a critical 
coupling, as in Eq.~(\ref{eq:critical_coupling_subtraction}),
and the leading physics
has {\em asymptotic scale invariance\/}---this 
applies to nonextremal metrics in the near-horizon expansion.

\subsection{Near-Horizon
WKB Framework:  Multidimensional Case}
\label{sec:IWKB_n-h_multidimensional}

The multidimensional equation~(\ref{eq:Schrodinger_equation}) 
describes the full-fledged spatial dependence of the modes.
The near-horizon expansion 
of the Laplace-Beltrami operator
for the metric~(\ref{eq:RN_metric}),
\begin{equation}
\check{\Delta}_{(\gamma)} 
=
  \frac{1}{ \sqrt{\gamma} } \partial_{j} 
\left[ 
\sqrt{\gamma} 
 \, \gamma^{jk} \, \partial_{k} 
\right]
\stackrel{(\mathcal H)}{\sim}
f'_{+}
\,
\left(
x \, \partial_{x}^{2} + \frac{1}{2} \,
\partial_{x}
\right)
\; ,
\end{equation}
implies that $p=-1$ 
(or ``$\check{\Delta}_{(\gamma)} \propto x^{-1}$'');
therefore, 
if $A(x) \propto x^{s} $, 
then
\begin{equation}
Q (x)
=
\frac{ \check{\Delta}_{(\gamma)} A (x)}{ A(x) }
\stackrel{(\mathcal H)}{\sim}
f'_{+}
\, s \, \left( s- \frac{1}{2} \right)
\,
\, x^{-1}
\;  .
\label{eq:quantum_potential_leading_explicit}
\end{equation}
Clearly, for the 
nonextremal metrics, the leading scale dimension
of $\mathfrak{I} (x)
\stackrel{(\mathcal H)}{\propto}  1/x$
is equal to 1, thus
showing that this is a marginally singular case:
the near-horizon physics
exhibits {\em SO(2,1) conformal invariance\/}.
Accordingly, the semiclassical function
$\tilde{ \mathfrak{I} } (x)$ is given by Eq.~(\ref{eq:critical_coupling_subtraction})
with a critical coupling
\begin{equation}
 c^{(*)} =
f'_{+}
\, s \, \left(  \frac{1}{2} - s
\right)
\; .
\label{eq:critical_coupling_multidimensional}
\end{equation}

In addition,
the value of the parameter $s$ for the multidimensional
case can be determined 
from  the continuity equation~(\ref{eq:probability_conservation}),
which yields the leading order of the amplitude through
\begin{equation}
\frac{ \partial}{ \partial x} 
\left( 
\frac{ \gamma^{xx} }{ \sqrt{f} } \, A^{2} k_{x} 
\right)
\stackrel{(\mathcal H)}{\propto}
\frac{ \partial}{ \partial x} 
\left( 
 \sqrt{x}  \, A^{2} k_{x}
\right)
\stackrel{(\mathcal H)}{\sim}
0
\; .
\label{eq:probability_conservation_expansion}
\end{equation}
Therefore,
$\hat{k}_{x} 
\equiv 
\sqrt{ \gamma^{xx} } \, k_{x} 
=
\sqrt{f} \, k_{x} 
\stackrel{(\mathcal H)}{\propto}
\sqrt{x} \, k_{x}
$
gives the amplitude scaling
\begin{equation}
A (x) 
\stackrel{(\mathcal H)}{\propto}
 (\hat{k}_{x})^{-1/2}
\stackrel{(\mathcal H)}{\propto}
\left[
\tilde{ \mathfrak{I} } 
(x)
\right]^{-1/4} 
\stackrel{(\mathcal H)}{\propto}
 x^{1/4}
\; ,
\label{eq:multidimensional_amplitude_scale}
\end{equation}
where,
from Eq.~(\ref{eq:eta_definition}),
 $\tilde{ \mathfrak{I} } (x) 
\stackrel{(\mathcal H)}{\sim}
\gamma^{xx}
\,
(k_{x})^{2}
=
(\hat{k}_{x})^{2}
\stackrel{(\mathcal H)}{\propto}
1/x$
($p=q=-1$).
In particular,
Eq.~(\ref{eq:multidimensional_amplitude_scale})
implies that  $s=1/4$;
as a result, from
Eq.~(\ref{eq:quantum_potential_leading_explicit}),
the leading ``extra term'' becomes
\begin{equation}
Q (x)
\stackrel{(\mathcal H)}{\sim}
- f'_{+}   \,
\frac{1}{16}\, 
\frac{1}{x}
\stackrel{(\mathcal H)}{\sim}
 - f(x) \,
\frac{1}{16}\, 
\frac{1}{x^{2} }
\; .
\label{eq:quantum_potential_1}
\end{equation}

Finally, from the near-horizon expansion of 
Eq.~(\ref{eq:Klein_Gordon_normal_QMreduced_invariant_NH}),
and Eqs.~(\ref{eq:n-h_diff_eq_for_eta}) and 
(\ref{eq:quantum_potential_1}),
\begin{equation}
\tilde{k}
\equiv 
\sqrt{
\frac{
\tilde{ \mathfrak{I} } (x)
}{f(x) }
}
 \stackrel{(\mathcal H)}{\sim} 
\sqrt{
\frac{
\mathfrak{I} (x) + Q[\mathfrak{I}] (x)
}{f(x) }
}
\stackrel{(\mathcal H)}{\sim} 
\frac{\Theta }{ x }
\; ,
\label{eq:WKB_physical_radial_wavenumber_n-h_expansion}
\end{equation}
which defines an improved wave number.
Then, the leading form of $\tilde{\mathfrak{I}} (x) $
yields the chain of relations,
$
\tilde{k} 
 \stackrel{(\mathcal H)}{\sim} 
k_{r}
 \stackrel{(\mathcal H)}{\sim}  
k_{\rm conf} (x)
$,
which reduce to the 
{\em conformal wave number\/} 
\begin{equation}
k_{\rm conf} (x)
\equiv \frac{\Theta}{ x }
\;  .
\label{eq:conformal_wavenumber_def}
\end{equation}

In conclusion,
this calculation
shows that:
(i)
the leading covariant momentum component
is radial;
(ii)
 $k_{\rm conf} (x)$
embodies the improved WKB features
of conformal quantum mechanics;
and (iii)
 $k_{\rm conf} (x)$ is the correct input for the phase-space algorithms 
of Sec.~\ref{sec:phase_space_semiclassical_methods}.

\subsection{Near-Horizon WKB Framework: 
Reduced Radial Case}
\label{sec:IWKB_n-h_reduced-radial}

Equation~(\ref{eq:Klein_Gordon_normal_radial})
was derived through the sequence of exact Liouville transformations;
in turn, this equation can be solved within the semiclassical
approximation, with
\begin{equation}
{\Delta}_{(1D)} 
\equiv
 \partial_{x}^{2}
\,  
\label{eq:1D_LaplaceBeltrami}
\end{equation}
applied to the formalism of subsection~\ref{sec:WKB_Expansion}.
 The original radial part of the Laplacian
also includes the prefactor $f(r)$; however,
in the sequence of transformations leading 
to Eq.~(\ref{eq:Klein_Gordon_normal_radial}),
$f(r)$ was scaled away.
As a result, the leading scaling of 
Eq.~(\ref{eq:1D_LaplaceBeltrami}) is
now given by
\begin{equation}
Q (x)
= \frac{  A'' (x)}{ A(x) }
=
s \, \left( s- 1 \right)
\,
\, x^{-2}
\;  ,
\label{eq:quantum_potential_leading_radial1D}
\end{equation}
i.e.,
``${\Delta}_{(1D)} \propto x^{-2}$.''
Moreover, the near-horizon leading form of 
Eq.~(\ref{eq:Klein_Gordon_normal_radial})
becomes 
\begin{equation}
{\Delta}_{(1D)} u(x)
 + 
\left[
\frac{ 
\Theta^{2}
+ 1/4 }{x^{2}}
\right]
\,
u (x)
=
0
\;  ,
\label{eq:CQM_equation}
\end{equation}  
which corresponds to 
the effective conformal interaction~(\ref{eq:conformal_interaction})
and implies that
\begin{equation}
\mathfrak{I} (x)
\stackrel{(\mathcal H)}{\sim}
\left( \Theta^{2} + \frac{1}{4} \right)
\,
\frac{1}{x^{2} }
\; .
\label{eq:n-h_leading_I_radial}
\end{equation}
Accordingly, in the radial setup of
the generalized WKB framework, the scale dimensions of 
Eqs.~(\ref{eq:1D_LaplaceBeltrami}),
(\ref{eq:quantum_potential_leading_radial1D}),
and
(\ref{eq:n-h_leading_I_radial}) are equal to 2
for the nonextremal metrics:
the near-horizon physics is marginally singular, 
with the scale symmetry of conformal quantum mechanics.

In addition, the parameter $s$ is determined 
from the leading-order continuity equation,
\begin{equation}
\frac{ \partial}{ \partial x} 
\left(
 A^{2} k_{x} 
\right)
\stackrel{(\mathcal H)}{\sim}
\frac{ \partial}{ \partial x} 
\left\{ 
   A^{2} 
\left[
\tilde{ \mathfrak{I} } (x)
\right]^{1/2} 
\right\}
\stackrel{(\mathcal H)}{\sim}
0
\; ,
\label{eq:probability_conservation_expansion_radial}
\end{equation}
where
$k_{x} 
\stackrel{(\mathcal H)}{\sim}
\left[
\tilde{ \mathfrak{I} } (x)
\right]^{1/2} 
$ 
for the one-dimensional analogue
of Eq.~(\ref{eq:eta_definition}).
In turn, 
\begin{equation}
A (x)
\stackrel{(\mathcal H)}{\propto}
\left[
\tilde{ \mathfrak{I} } 
(x)
\right]^{-1/4} 
\stackrel{(\mathcal H)}{\propto}
 x^{1/2}
\; ,
\label{eq:radial_amplitude_scale}
\end{equation}
because $\tilde{ \mathfrak{I} } (x) 
\propto 1/x^{2}$ for the reduced radial case
(as $p=q=-2$).
In particular,
Eq.~(\ref{eq:radial_amplitude_scale})
 shows that  $s=1/2$ and yields the critical coupling
$ c^{(*)} = 1/4$,
as Eq.~(\ref{eq:quantum_potential_leading_radial1D})
turns into
\begin{equation}
Q (x) 
\stackrel{(\mathcal H)}{\sim}
-
\frac{1}{ 4 x^{2}}
\; .
\label{eq:WKB_error_conformal}
\end{equation}
Thus, the one-dimensional analogue of 
Eqs.~(\ref{eq:eta_definition})
and (\ref{eq:n-h_diff_eq_for_eta})
yields the conformal behavior~(\ref{eq:conformal_wavenumber_def}),
\begin{equation}
 k_{\alpha_{l,D}} (r) 
\equiv 
\sqrt{
\tilde{ \mathfrak{I} }   (r; \omega, \alpha_{l,D}  ) 
}
\stackrel{(\mathcal H)}{\sim} 
k_{\rm conf} (x)
\; .
\label{eq:WKB_physical_radial_wavenumber_radial-eq}
\end{equation}

In conclusion,
Eq.~(\ref{eq:WKB_physical_radial_wavenumber_radial-eq})
provides the wave number 
for the WKB wave functions
\begin{equation}
 u_{\pm} (r) 
=
\left[
k_{\alpha_{l,D} } (r)
\right]^{-1/2} 
 \exp \left[
\pm i \int^{r} 
k_{\alpha_{l,D}} (r') dr'
\right]
\; .
\label{eq:WKB_wave_functions_general}
\end{equation}
Even though the variables 
$\tilde{k}$ of Eq.~(\ref{eq:WKB_physical_radial_wavenumber_n-h_expansion})
and 
$k_{\alpha_{l,D}} (r)$
of Eq.~(\ref{eq:WKB_physical_radial_wavenumber_radial-eq})
are different, their near-horizon leading contributions reduce 
to the same conformal value~(\ref{eq:conformal_wavenumber_def}).
Moreover, 
Eq.~(\ref{eq:scalar_field_in_BH_background_5_invariant_NH})
implies the competition 
of the angular momenta with
 $k_{\rm conf} (x) $ in the form
\begin{equation}
 k_{\alpha_{l,D}} (r= r_{+}+x; \Theta,
\alpha_{l,D}
) 
\stackrel{(\mathcal H)}{\sim}
k_{\rm conf} (x) \,
\sqrt{ 1 
- 
\frac{ 
\alpha_{l,D} \, x}{ f'_{+} \, r_{+}^{2} \Theta^{2}
    }
}
\; .
\label{eq:WKB_wavenumber_nh_explicit}
\end{equation}

\section{Phase-Space Methods for Quantum Mechanics
and Quantum Field Theory in Curved Spacetime}
\label{sec:phase_space_semiclassical_methods}


The main goal of this section is to derive
phase-space expressions---compatible with the 
improved WKB approach---for 
 the cumulative number of modes or spectral function
\begin{equation}
N(\omega)
=
\sum_{
\stackrel{s }{\omega_{s} \leq \, \omega}
}
1
\; .
\label{eq:cumulative_number_of_states_def}
\end{equation}
For a monotonic increasing operator~\cite{Sturm}
 $ -\hat{\mathcal H}_{\rm eff} (\omega ) $,
Eq.~(\ref{eq:cumulative_number_of_states_def})
is equivalent to
\begin{equation}
N(\omega)
= 
{\rm Tr} \, 
\biggl[
\theta
\biggl( 
-
\hat{\mathcal H}_{\rm eff} 
(\omega)
\biggr)
\biggr]
=
\sum_{ s}
\,
\theta
\biggl( 
-
\left[\hat{\mathcal H}_{\rm eff} (\omega)\right]_{s}
\biggr)
\; ,
\label{eq:cumulative_number_of_states}
\end{equation}
in which $\theta ( z  )$ stands for the Heaviside function
and the formal trace
is defined in the Hilbert space 
spanned by the 
basis of modes $\phi_{s}(\vec{x})$.

\subsection{Phase-Space Method: Generic Techniques}
\label{sec:phase_space}

 For the statistical mechanics
of a quantum-mechanical system in {\em curved space\/},
the semiclassical counterpart of
 Eq.~(\ref{eq:cumulative_number_of_states}) 
is derived by counting the number
of phase-space cells 
$d \Gamma  /(2\pi)^{d}$
enclosed within a given 
$\omega$-parametrized surface
${\mathcal H}_{\rm eff}( \vec{x},\vec{p}; \omega) = 0$; 
this is computed with the Liouville measure
in local Darboux coordinates~\cite{symplectic}
$
d \Gamma  
=
dx^{1} \wedge \ldots \wedge 
dx^{d} 
\,
\wedge 
\,
dp_{1}  \wedge 
\ldots  \wedge 
dp_{d} 
$---with the shorthand
$
d \Gamma  
=
 d^{d} x
\,
d^{d} p$,
in terms of the covariant momentum components.
Then, for a classical Hamiltonian 
${\mathcal H}_{\rm eff}( \vec{x},\vec{p}) $,
with configuration-space metric $\gamma_{ij} (\vec{x})$,
\begin{equation}
N ( {\omega})
\approx
\int
\frac{ d \Gamma }{ (2 \pi )^{d} }
\;
\theta 
\biggl(
 - {\mathcal H}_{\rm eff}( \vec{x},\vec{p}; \omega) 
\biggr)
=
\frac{ 1 }{ (2 \pi )^{d} }
\,
\int 
d^{d} x
\,
\sqrt{\gamma}
\,
\int\limits_{ 
{\mathcal H}_{\rm eff}( \vec{x},\vec{p}; \omega) 
\leq 
0
}
 d^{d} p
\,
\frac{1}{\sqrt{\gamma}}
\; 
\label{eq:phase_space_spectral_number}
\end{equation}
(where the symbol $\approx$ denotes
the semiclassical approximation {\em before a hierarchical expansion\/}).

For the analysis of the black-hole problem 
of Eq.~(\ref{eq:Klein_Gordon_normal_spatial}),
the momentum dependence of the effective Hamiltonian 
$\hat{\mathcal H}_{\rm eff}(\omega)$
is merely quadratic and
two distinct ways of evaluating
Eq.~(\ref{eq:phase_space_spectral_number}) are possible:
(i) the {\em multidimensional\/} approach
and 
(ii) the  {\em  reduced radial\/} approach.

The multidimensional approach starts by integrating out 
{\em all\/} the generalized momenta:
\begin{equation}
N ( {\omega})
\approx
\frac{ \Omega_{(d-1)}}{ d \, ( 2 \pi)^{d} }
\,
\int 
 d V_{(d)} 
\,
\,
\| \vec{k}  (\vec{x}) \|^{d}
\; ,
\label{eq:phase_space_spectral_number_explicit_fully_multidim}
\end{equation}
where
$ d V_{(d)} 
= d^{d} x \, 
\sqrt{\gamma}$
is the $d$-dimensional spatial volume element
and $\| \vec{k}  (\vec{x}) \|
\equiv 
\| \vec{p}  (\vec{x}) \|
=
\sqrt{ \gamma^{jh} ( \vec{x} )  
\,
p_{j} ( \vec{x}  ) p_{h}  ( \vec{x} ) }
$
(with $\vec{k } 
\equiv \vec{p }$).
In addition,
in the presence of a hierarchical expansion
[from Eqs.~(\ref{eq:eta_definition})
and (\ref{eq:n-h_diff_eq_for_eta})],
$
\| \vec{k}  (\vec{x}) \|
\sim
\biggl\{
\mathfrak{I} (\vec{x}) 
+ 
Q [\mathfrak{I}] (\vec{x}) 
\biggr\}^{1/2}
$,
with $\approx$ replaced by $\sim$.
Moreover,
when the potential 
is spherically symmetric: $ \tilde{\mathfrak{I}} ( \vec{x} )
= \tilde{\mathfrak{I}} (r)$,
Eq.~(\ref{eq:phase_space_spectral_number_explicit_fully_multidim})
becomes
\begin{equation}
N ( {\omega})
\approx
\frac{ [\Omega_{(d-1)}]^{2} }{ d \, ( 2 \pi)^{d} }
\,
\int 
dr 
\,
 [ \gamma_{rr} ]^{-(d-1)/2}
\,
r^{d-1} 
\,
\left[
\tilde{k}(r)
\right]^{d}
\; ,
\label{eq:phase_space_spectral_number_polar_coords_explicit_alternative}
\end{equation}
where
\begin{equation}
\tilde{k}( r ) 
= [ \gamma_{rr} ]^{1/2} 
\, 
\| \vec{k}  (r) \|
\;  .
\label{eq:multidim_covariant__semiclassical_wave_number}
\end{equation}

In the radial approach, for a spherically symmetric Hamiltonian,
 Eq.~(\ref{eq:phase_space_spectral_number})
turns into a radial integral in configuration
space and an integral over the angular momenta;
this is accomplished by a four-step
method.
First, 
from the polar coordinates
$\vec{x} \equiv ( r, \theta^{1}, \cdots, \theta^{d-1})$,
the conjugate momenta 
$\vec{p} \equiv ( p_{r}, \ell_{1}, \cdots, \ell_{d-1})$
satisfy
$\gamma^{jk} p_{j} p_{k}
= 
\gamma^{rr} 
 p_{r}^{2}
+ \ell^{2}/r^{2}
$, 
where $\ell_{a}$ are angular momenta
(with $a=1, \cdots, d-1$)
and
$ \ell^{2} = \ell^{a} \ell_{a}$.
Second, the radial momentum can be integrated out
with
$
\int_{-\infty}^{\infty}    
d p_{r}
\,
\theta 
\biggl(
 \tilde{\mathfrak{I}} ( r )
-
\gamma^{rr} 
 p_{r}^{2}
- \alpha_{l}/r^{2}
\biggr)
=
2 \, \tilde{k} (r;\alpha_{l} )
$,
where $\alpha_{l} = \ell^{2}$
and
\begin{equation}
\tilde{k} (r;\alpha_{l} )
\equiv 
k_{\alpha_{l}, D} (r)
=
\left( \gamma^{rr}  \right)^{-1/2}
\,
\sqrt{ 
 \tilde{\mathfrak{I}} ( \vec{x} )
 - 
\frac{\alpha_{l} }{  r^{2}  } 
}
 = 
\sqrt{ 
\left[\tilde{k}( r )  \right]^{2}
 - 
  \gamma_{rr}
\, 
\frac{\alpha_{l} }{  r^{2} } } 
\; ,
\label{eq:radial_semiclassical_wave_number}
\end{equation}
with $\tilde{k}( r ) $ defined by
Eq.~(\ref{eq:multidim_covariant__semiclassical_wave_number}).
Third, 
the angular-momentum dependence
is kept through 
$
\alpha
\equiv
\alpha_{l}
$ and with the
use of 
$
\int 
d^{d-1} \ell/\sqrt{\sigma} 
=
\Omega_{(d-2)}
\int d \alpha
\,
\alpha^{(d-3)/2}/2
$,
where 
$\sigma_{ab}$ is the $S^{d-1}$ metric
associated with
$\Omega \equiv
\{ \theta^{a} \}$ ($a=1, \ldots , d-1$).
Finally, integration of the angular variables
$d \Omega_{(d-1)} \equiv
d^{d-1} \theta \, \sqrt{\sigma}$
yields the solid angle 
$\Omega_{(d-1)}$.
Thus, 
the spectral function becomes
\begin{equation}
N ( {\omega})
\approx
\frac{
\Omega_{(d-1)}
\Omega_{(d-2)}}{  ( 2 \pi)^{d} }
\int d \alpha 
\alpha^{(d-3)/2}
\int_{\mathcal I}
 dr 
\,
\tilde{k} (r;\alpha )
\; ,
\label{eq:phase_space_spectral_number_polar_coords_explicit}
\end{equation}
where the  interval
${\mathcal I}$ is bounded by the classical turning points;
in the presence of a hierarchical expansion,
Eq.~(\ref{eq:phase_space_spectral_number_polar_coords_explicit})
requires the use of {\em improved wave numbers\/}.
Equivalently, Eq.~(\ref{eq:phase_space_spectral_number_polar_coords_explicit})
has been shown to follow from the one-dimensional 
Sturm oscillation theorems~\cite{BH_thermo_CQM}.

\subsection{Phase-Space Method: 
Quantum Field Theory in Curved Spacetime \& Near-Horizon Physics}
\label{sec:semiclassical_field_theory}

We now turn to the specific computation of the 
spectral number function $N(\omega)$ 
corresponding to our quantum field theory in {\em curved spacetime\/}.
The starting point is the spatially reduced Klein-Gordon 
equation~(\ref{eq:Klein_Gordon_normal_spatial}).
Its classical limit involves a simple Hamiltonian formulation
with the modification~(\ref{eq:WKB_quantum_potential_correction})
at the level of the effective potential.
Consequently,
\begin{equation}
N (\omega)
\approx
\int 
d^{d} x
\,
\int
\frac{ d^{d} p}{ (2 \pi )^{d} }
\;
\theta 
 \biggl(
\mathfrak{I}_{(0)} (r;\omega)
+
\mathfrak{I}_{(1)} (r)
+
Q(r)
-
\gamma^{jk} (\vec{x})
\,
p_{j} p_{k}
 \biggr)
\; ,
\label{eq:phase_space_spectral_number_Klein_Gordon}
\end{equation}
where the ``quantum potential'' $Q(r)$ 
is required for the near-horizon
expansion of nonextremal metrics, and
the approaches of the previous
subsection can be applied.

In the multidimensional approach,
Eq.~(\ref{eq:phase_space_spectral_number_Klein_Gordon})
leads to the counterpart of
Eqs.~(\ref{eq:phase_space_spectral_number_explicit_fully_multidim})
and
(\ref{eq:phase_space_spectral_number_polar_coords_explicit_alternative}),
with 
$\gamma_{rr} (r) = 1/f(r)$;
 in the near-horizon limit,
from Eq.~(\ref{eq:WKB_physical_radial_wavenumber_n-h_expansion}),
\begin{equation}
N(\omega) 
\stackrel{(\mathcal H)}{\sim}
\frac{1}{d \, 2^{d-2} \, [\Gamma (d/2)]^{2} }
\,
\int 
dx 
\,
r_{+}^{d-1} 
\,
\underbrace{
\left[
f'_{+} \, x
\right]^{(d-1)/2}
}_{\rm angular \; contribution}
\;
\underbrace{
\left[
k_{\rm conf} (x) 
\right]^{d}
}_{\rm conformal \; interaction}
\; ,
\label{eq:phase_space_spectral_number_polar_coords_explicit_QFT_n-h}
\end{equation}
which displays a competition 
of the conformal wave number 
$k_{\rm conf} (x)$ 
with the angular-momentum  factors 
$
\left[
f'_{+} 
\,
x  
\right]^{(d-1)/2}
$.
These factors reduce the degree of divergence of the integral,
but the ensuing singular behavior can be ultimately
attributed to the ultraviolet 
singularity of conformal quantum mechanics~\cite{camblong}.
As a final step, from 
Eqs.~(\ref{eq:conformal_wavenumber_def})
and (\ref{eq:phase_space_spectral_number_polar_coords_explicit_QFT_n-h}),
 \begin{equation}
 N(\omega)
\stackrel{(\mathcal H)}{\sim}
\frac{1}{\pi
   \Gamma (d-1) }
\,
B \left( \frac{d-1}{2} , \frac{3}{2} \right) \,
\Theta^{d} 
\,
\left[ f'_{+} \, r_{+}^{2}  \right]^{ (d-1)/2 }
\,
\lim_{a \rightarrow 0}
\int_{a}^{x_{1}} \frac{dx}{x^{(d+1)/2}}
\; ,
\label{eq:WKB_number_of_states_2}
\end{equation}
where $a$ is a radial cutoff and
$B(p,q)$ is the beta function, while $x_{1}$ 
is an arbitrary upper limit (with a scale of the order of $r_{+}$).
This cutoff and the 
associated renormalization of Eq.~(\ref{eq:WKB_number_of_states_2})
are discussed in the next section
and analyzed in Ref.~\cite{BH_thermo_CQM}.

In a similar manner, for the reduced radial problem,
Eq.~(\ref{eq:WKB_wavenumber_nh_explicit})
turns Eq.~(\ref{eq:phase_space_spectral_number_polar_coords_explicit})
into
\begin{equation}
 N(\omega) 
\stackrel{(\mathcal H)}{\sim}
\frac{1}{\pi \, \Gamma (d-1) }
\int_{0}^{\alpha_{\rm max}} 
d \alpha \,
\alpha^{d/2- 3/2} \, 
\int_{\mathcal I} 
dx
\,
k_{\rm conf} (x)
\,
\sqrt{ 1 
- 
\frac{ 
\alpha \, x}{ f'_{+} \,  r_{+}^{2} \Theta^{2}
    }
}
\; ,
\label{eq:semiclassical_number_of_states}
\end{equation}
where $\alpha_{\rm max} =
\alpha_{\rm max} (a) = \Theta^{2} f_{+}' r_{+}^{2}/a$ is the
angular-momentum cutoff arising from the passage of the right turning point
through $r=a$.
Finally, reversing the order of integration and 
using a beta-function identity,  Eq.~(\ref{eq:WKB_number_of_states_2})
follows again.
This shows the equivalence of the reduced radial
and multidimensional approaches.

\section{Bekenstein-Hawking Entropy 
from the Near-Horizon Expansion: Conclusions}
\label{sec:BHEntropy_conclusions}

In this paper we have illustrated the use of
improved semiclassical techniques 
for the computation of spectral functions and derived the 
corresponding near-horizon expansions, with the central
 result~(\ref{eq:WKB_number_of_states_2})
being independent of the 
semiclassical procedure involved.

Unfortunately, 
as it stands,
Eq.~(\ref{eq:WKB_number_of_states_2}) 
appears to be divergent 
when the lower limit $a$ approaches zero.
This singularity
can be traced to the scale invariance
of the effective conformal interaction and
is inherited by the thermodynamic observables.
The cutoff $a$ serves as a regulator
and leads to the renormalization of the theory,
which can be implemented geometrically by absorbing 
the coordinate assignment $a$ into a 
distance or ``elevation''
\begin{equation}
h_{D} =
\int_{ r_{+}}^{r_{+}+a} ds
\stackrel{(\mathcal H)}{\sim}
\frac{ 2\sqrt{a} }{ \sqrt{ f'_{+} }  }
\label{eq:brick_wall_geometrical_elevation_regularized}
\end{equation}
from the horizon.
As shown in Ref.~\cite{BH_thermo_CQM},
the various contributions to the entropy can be 
organized into those factors that are purely conformal and 
those arising from the  angular momentum:
their interplay leads to the familiar Bekenstein-Hawking entropy
$S_{\rm BH}= {\mathcal A} /4$,
which is a $(D-2)$-dimensional ``hypersurface'' 
feature, induced by the horizon.
Furthermore, this result relies on the purely conformal characterization
of the Hawking temperature~\cite{padmanabhan_HawkingT,BH_thermo_CQM}
needed in the statistical-mechanical
calculations. Moreover, 
this procedure shows that the ``new physics'' of a full-fledged
quantum gravitational theory arises from within an
invariant distance of the order of the Planck length. 

Despite its appealing features discussed above,
the regularization procedure based on the 
brick-wall model leaves a number of
unanswered questions.
First, the computation leading
to the Bekenstein-Hawking result,
with the correct numerical prefactor of $1/4$, involves
 a fine tuning of the cutoff~\cite{BH_thermo_reviews,frolov,thooft:85}.
This poses a problem:
in this method of calculation, the numerical prefactor appears to
depend upon the number $Z$ of species of particles,
rather than being a $Z$-independent value of 1/4;
for example, in the case of $Z$ scalar fields, the required
fine tuning involved in
Eq.~(\ref{eq:brick_wall_geometrical_elevation_regularized})
would lead to a brick-wall elevation with the species dependence
\begin{equation}
 h_{D}
=
\frac{1}{2}
\left[
 D \zeta (D) \Gamma (D/2-1) \pi^{1-3D/2} \, Z
\right]^{1/(D-2)}
\; .
\label{eq:brick_wall_geometrical_elevation}
\end{equation}
Another paradoxical feature of
the entropy computed by a brick-wall method is that 
it can be absorbed by a renormalization of Newton's gravitational constant 
$G_{N}$, as shown in 
Refs.~\cite{susskind_uglum,renormalization_Newton_confirmed}.
However,
in this light, it is possible that the species problem associated 
with the entropy prefactor may be compensated by a corresponding 
$Z$-dependent renormalization
of Newton's 
constant~\cite{frolov,renormalization_BW_miscellaneous,BH_confirmed_Boulware}.

In summary,
we have established that
the procedure that singles out the leading conformal behavior also
provides a systematic application of {\em semiclassical methods\/}.
The robust nature of this framework
and the asymptotically exact semiclassical description
of conformal quantum mechanics 
are somewhat surprising, given the presence of a coordinate
singularity.
Remarkably, these techniques: (i)
 converge towards a {\em unique\/}
result driven by the near-horizon expansion,
the Bekenstein-Hawking entropy;
(ii) point to the horizon degrees of freedom that determine the 
thermodynamics.
The fact that this universality is driven by the near-horizon
symmetry is intriguing, but its deeper geometrical meaning is
not well understood.
However, the robustness and simplicity of these properties suggest
their possible origin from an even more fundamental principle of nature.

\acknowledgments{
This research was supported 
by the National Science Foundation under Grant 
No.\ 0308300 
(H.E.C.) and under Grant No.\ 0308435 (C.R.O.), and
by the University of San Francisco Faculty Development Fund
(H.E.C.).
We also thank 
Professor Cliff Burgess
for stimulating discussions and
Dr.\ Stanley Nel for generous travel support 
that facilitated the conception of this project.
}

\end{document}